\documentclass[graybox,natbib,nosecnum]{svmult}
\bibpunct{(}{)}{;}{a}{}{,} 

\pdfoutput=1   

\usepackage{mathptmx}       
\usepackage{helvet}         
\usepackage{courier}        
\usepackage{type1cm}        

\usepackage{makeidx}         
\usepackage{graphicx}        
\usepackage{multicol}        
\usepackage[bottom]{footmisc}
\usepackage[normalem]{ulem}	
\usepackage{hyperref}  

\usepackage{soul}   

\makeindex             

\begin{document}

\title*{Metal-Depleted Brown Dwarfs}
\author{Nicolas Lodieu}
\institute{Instituto de Astrof\'\i sica de Canarias, C. Via Lactea S/N, E-38205 La Laguna, Tenerife, Spain, \email{nlodieu@iac.es}}
%
%
\maketitle

%
%
\abstract{
This chapter reviews our current knowledge of metal-poor ultracool dwarfs with spectral types
later than M7\@. The current census of M, L, and T subdwarfs is explored. The main colour trends 
of subdwarfs from the optical to the mid-infrared are described and their spectral features presented,
which led to a preliminary and tentative spectral classification subject to important changes in the
future when more of these metal-poor objects are discovered. Their multiplicity and the determination
of their physical parameters (effective temperature, gravity, metallicity, and 
mass) are discussed. Finally, some suggestions and future guidelines are proposed to
foster our knowledge on the oldest and coolest members of our Galaxy.
}

%
%
\section{Introduction}
\label{review_subBDs:intro}

M dwarfs represent the majority of stars in the solar neighbourhood \citep{kirkpatrick12} and in our 
Galaxy where the mass function peaks \citep[e.g.][]{chabrier03}. At lower masses, three new classes have
defined during the past two decades: the L dwarfs whose atmospheres are affected by dust
\citep{kirkpatrick99,martin99a}, the T dwarfs shaped by methane and water absorption bands
\citep{leggett00,geballe02,burgasser02,burgasser06a}, and the Y dwarfs with the potential presence
of ammonia at infrared wavelengths \citep{cushing11,kirkpatrick12}. The classification of L dwarfs is 
mostly morphological but the large variety of sources discovered in optical and infrared large-scale 
surveys triggered a preliminary spectral scheme incorporating a new parameter: gravity (i.e.\ ages)
as proposed by two independent teams \citep{cruz09,allers07,allers13}. However, the spectral 
classification of metal-poor
L dwarfs remains in its infancy due to the small sample existing in the literature. Nonetheless,
recent discoveries offered new hints to elaborate a tentative spectral sequence.

Metal-poor dwarfs belong to the spectral class VI in the Morgan-Keenan scheme \citep{morgan43}. They
are also known as subdwarfs and often abbreviated ``sd'' \citep{joy47a,gizis97a,lepine07c}. They usually 
exhibit bluer optical and infrared colours than their solar-like analogues \citep{lodieu17a}
and show distinct spectral features such as the weakening of TiO bands (i.e.\ less TiO opacity
implying more flux radiated from deeper and hotter layers of the atmosphere), strengthening of CaH bands,
and strong collision-induced hydrogen absorption beyond 1$\mu$m \citep{gizis97a,lepine07c}. They typically have
high proper motions and large radial velocities translating into space motions compatible with membership
to the thick disk and halo \citep{schmidt75a}. This population of metal-poor dwarfs is important for
several reasons. Firstly, they represent key tracers of the history of our Galaxy because they are very
old. Secondly, the knowledge of their physical parameters will impact on the study of globular clusters
whose main populations are metal-poor and old. Thirdly, the census of metal-poor stars and brown dwarfs
helps the determination of the luminosty and mass functions early on in the formation of our Galaxy
to gauge the impact of metallicity in star formation processes. Unfortunately, metal-poor stars are
not so numerous compared to their solar-like counterparts with only three subdwarfs of the 
$\sim$250 systems located within 10 pc ($\mu$\,Cas\,AB; Kapteyn's star; CF\,Uma).

This review will focus mainly on ultracool subdwarfs (UCSDs) with spectral types later than M7 and
metallicities (Fe/H) equal or less than $-$0.5 dex unless otherwise stated. For more massive 
subdwarfs, readers are referred to one of the section dedicated to M subdwarfs in the book of 
\citet*{reid05a}. The coolest L-type subdwarfs might be brown dwarfs but none of them has been
unambiguously proven to be substellar at the time of writing \citep{lodieu15a}. 
A few T-type metal-poor dwarfs have been announced as companions to bright stars with 
well-determined metallicities \citep{pinfield12,burningham13} 
but only one has a metallicity below $-$0.5 dex, WISE\,J20052038$+$5424339 \citep{mace13b}.

This review summarizes the techniques employed over the past decades to identify metal-poor low-mass stars 
and brown dwarfs, and describes the main spectral features leading to a preliminary and tentative 
classification scheme for UCSDs. The colours of UCSDs are mentioned and compared to those of nearby 
field M and L dwarfs. Our current knowledge on the multiplicity of UCSDs and the actual estimates of 
the physical parameters of the lowest mass metal-poor stars and brown dwarfs are also presented here. 
Finally, future needs are highlighted to characterise in more details the population of UCSDs and their 
physical parameters.

%
%
\section{Census of ultracool subdwarfs}
\label{review_subBDs:numbers}

Dedicated searches for metal-poor stars and brown dwarfs usually focus on proper motion surveys
to bias their final sample towards high velocity objects, thus halo stars \citep{schmidt75a}.
Most of the late-M subdwarfs have been identified in photographic plates
from the Digital Sky Survey and the SuperCOSMOS Sky Survey
\citep{gizis97a,gizis97b,gizis97c,schweitzer99,lepine03d,scholz04c,scholz04b,lodieu05b,lepine07c},
and more recent all-sky or large-scale surveys such as the Two Micron All Sky Survey 
\citep[2MASS;][]{burgasser06c,cushing09}, 
the Sloan Digital Sky Survey \citep[SDSS;][]{lepine08b,sivarani09,zhang13}, the UKIRT Infrared Deep 
Sky Survey \citep[UKIDSS;][]{lodieu12b,lodieu17a}, and the Wide field Infrared Survey Explorer
\citep[WISE;][]{kirkpatrick14,kirkpatrick16}. 

Over the past years, the number of L subdwarfs has grown rapidly and is now just slightly over 30\@. 
The first one was identified in 2MASS \citep{burgasser03b} followed by other discoveries in the 
same database \citep{burgasser04,cushing09},
SDSS \citep{sivarani09,schmidt10a,bowler10a}, WISE \citep{kirkpatrick14,kirkpatrick16}, and 
cross-correlations of various surveys \citep{lodieu10a,lodieu12b,lodieu17a,zhang17a,zhang17b}.

In the T dwarf regime, a few examples of metal-poor T dwarfs have been reported as companions to
brighter stars with well determined metallicities both in UKIDSS \citep{pinfield12,burningham13}
and WISE \citep{mace13b}. Nonetheless, their metallicities generally do not exceed $-$0.5 dex,
except in the case of WISE\,J200520$+$542433 with Fe/H\,=\,$-$0.64 dex \citep{mace13b}.

%
%
\section{Colours}
\label{review_subBDs:colours}

\begin{figure}
\includegraphics[width=0.35\linewidth,angle=0]{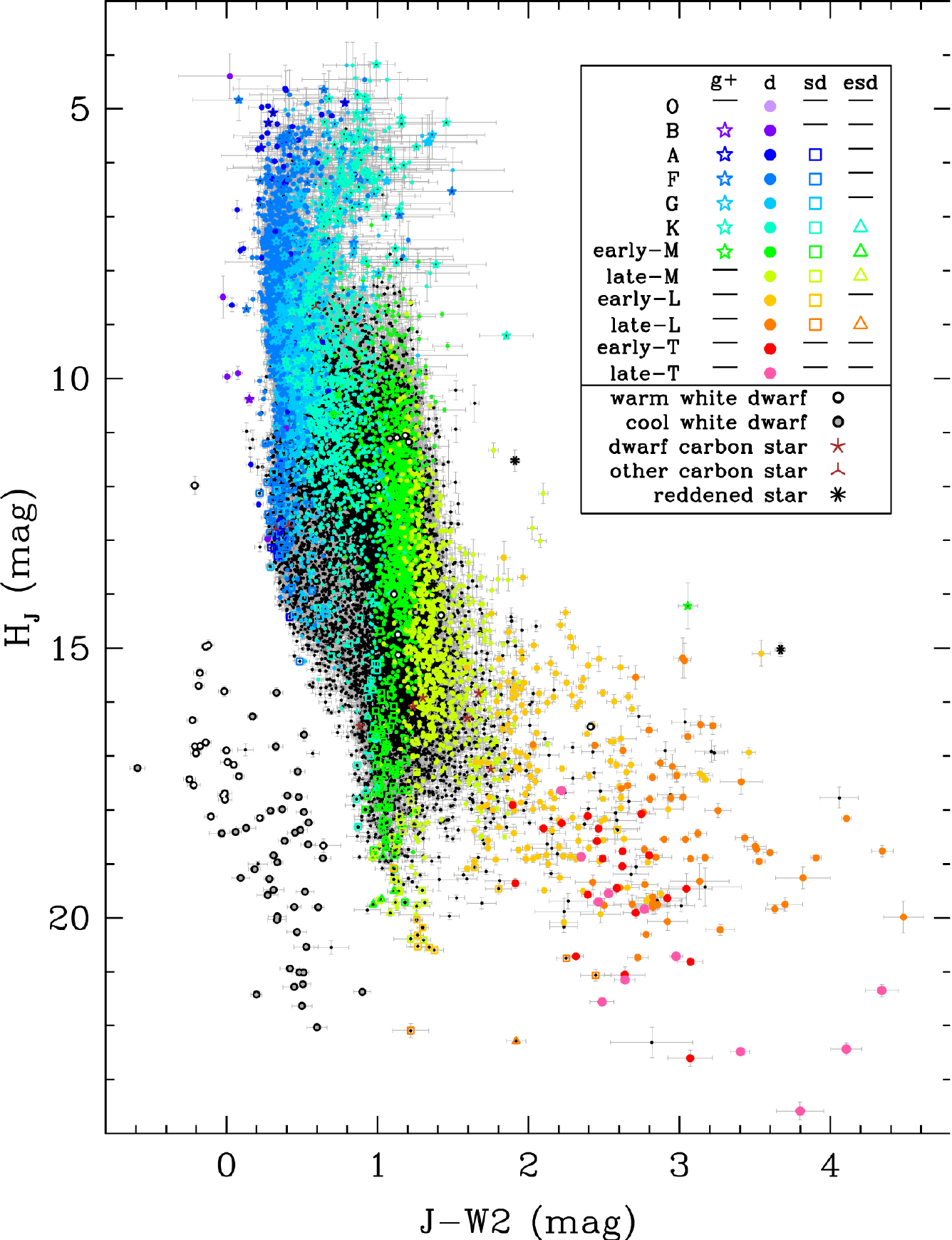}
\includegraphics[width=0.62\linewidth,angle=0]{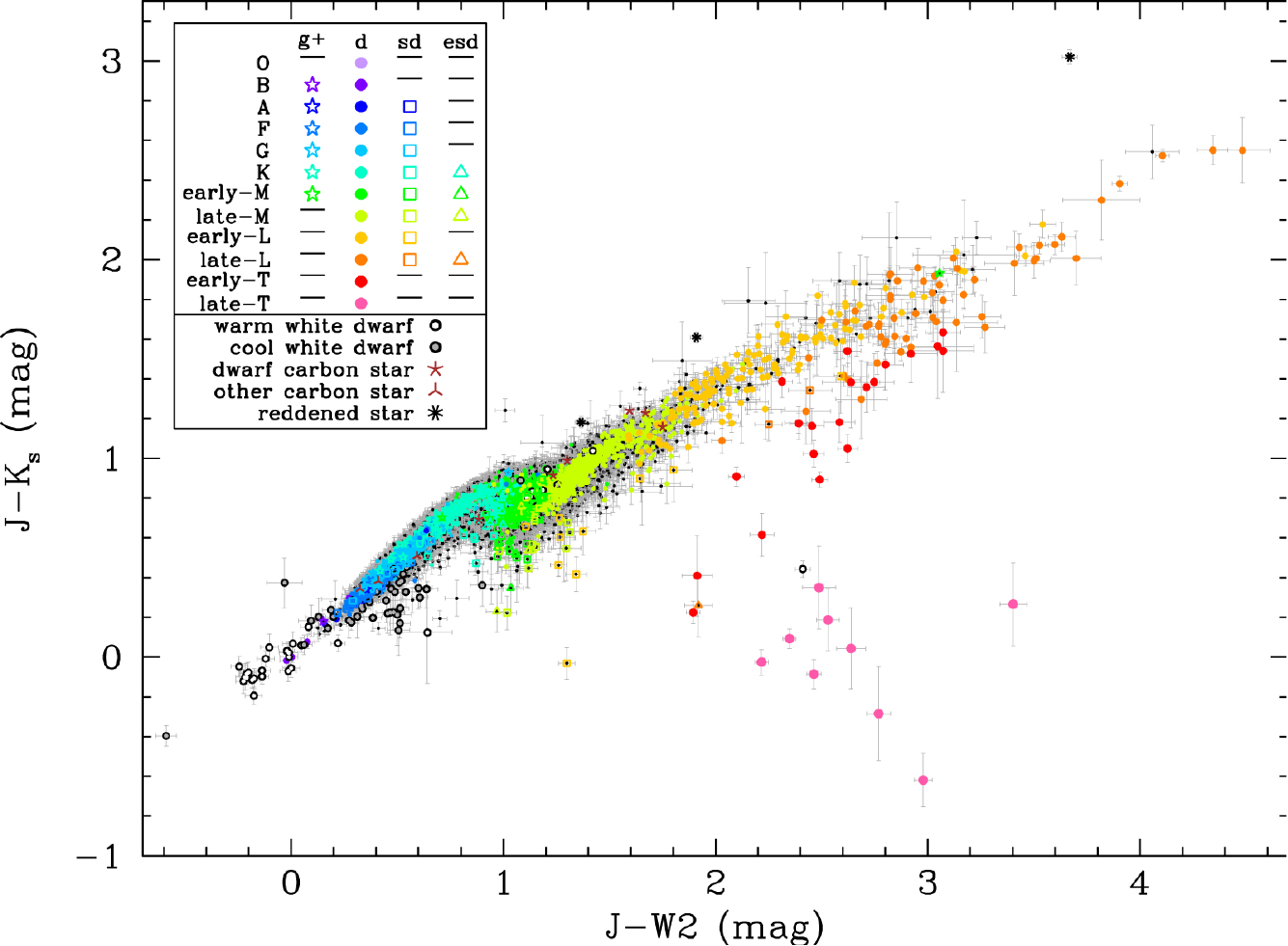}
\caption{Reduced proper motion (left) and ($J-w2$,$J-K_{s}$) colour-colour diagrams
for main-sequence stars, M, L, and T dwarfs of different metallicities.
Figures taken \citet{kirkpatrick16}.}
\label{review_subBDs:fig_diagrams}
\end{figure}

Due to the dearth of metals in their atmospheres, UCSDs tend to exhibit on average bluer colours than
their solar-type analogues (Fig.\ \ref{review_subBDs:fig_diagrams}). Moreover, they lie below solar-metallicity 
dwarfs in reduced proper motions (left panel in Fig.\ \ref{review_subBDs:fig_diagrams})
which constitute powerful tools to identify UCSDs \citep{lepine07c,lepine08b,kirkpatrick16,lodieu17a}.
This section summarises differences reported in the literature going from blue to red wavelengths.
These differences helped out increasing the numbers of UCSDs over the past years and represent key
indices for upcoming surveys like the Large Synoptic Survey Telescope \citep[LSST;][]{ivezic08a} 
and the Euclid mission \citep{mellier16a}.

\begin{itemize}
\item Their locus in a ($g-r$,$r-i$) colour-colour diagram is distinct from their solar-type counterparts.
Selecting sources with $g-r$\,$>$\,2 mag and $g-i$\,$>$\,3 mag will strongly bias the photometric selection
towards metal-poor M dwarfs \citep[Fig.\ 2 in][]{lepine08b}.
\item Their $r-z$ colours are bluer than field M dwarfs by at least 1 mag \citep[Fig.\ 2 in][]{lepine08b}.
\item The optical-to-infrared colours (e.g.\ $i-J$) of UCSDs are bluer as metallicity decreases, yielding 
a flattening of the far-red part of their optical spectra \citep{gizis97a,lepine07c,lodieu17a,zhang17a}.
\item Their infrared colours (e.g.\ $J-K$\,$<$\,0.7 mag) are much bluer than solar-type M dwarfs due to the
strong pressure-induced H$_{2}$ opacity beyond 1 micron \citep{lodieu17a,zhang17a}.
\item Subdwarfs of type M and L typically fall bluewards in near-infrared to mid-infrared colours
(e.g.\ $J-w1$, $J-w2$, $H-w2$) of their solar-metallicity counterparts due to the increasing influence 
of collision-induced hydrogen absorption \citep{kirkpatrick16,lodieu17a}.
\item Their mid-infrared colours look similar to those of their solar-type analogues although opposite trends
have been reported in the literature: red in \citet{kirkpatrick16} and blue in \citet{lodieu17a}.
\end{itemize}

%
%
%

%
%
\section{Spectral features}
\label{review_subBDs:spec_features}

The main spectral features indicative of low metallicity are the strengthening of the CaH bands and weakening
of the TiO bands around 620--740 nm and the effect of the collision-induced absorption beyond 1000 nm.
However, there are other features which can be employed to distinguish UCSDs from their solar-type cousins 
over the optical-infrared wavelength range \citep{gizis97a,lepine07c,burgasser07b,kirkpatrick16,zhang17a}, 
as enumerated below:

\begin{enumerate}
\item The CaH bands around 640--700 nm is stronger with lower metallicity but is also dependent on 
temperature in the 3600--3200\,K range.
\item The TiO bands at $\sim$720 nm, 780 nm, and 850 nm are weaker with lower metallicity. The bluest of
these bands, however, becomes more sensitive to temperature than metallicity for late-type M subdwarfs.
Below 3200\,K, the strength of the TiO band at 720 nm is not monotonic anymore with 
decreasing metallicity, making the classification of late-M and early-L subdwarfs based on spectral indices
more unreliable.
\item The VO band at 800 nm is a strong indicator of metallicity in L dwarfs: it weakens as Fe/H decreases.
\item The CO band at 2300 nm weakens as metallicity decreases and eventually disappears in extreme and ultra-subdwarfs.
\item The near-infrared flux beyond 1000 nm becomes more depressed with lower metallicity due to the
strong collision-induced H$_{2}$ absorption.
\end{enumerate}
\section{Spectral types}
\label{review_subBDs:spec_SpT}

\begin{figure}
\includegraphics[width=0.31\linewidth,angle=0]{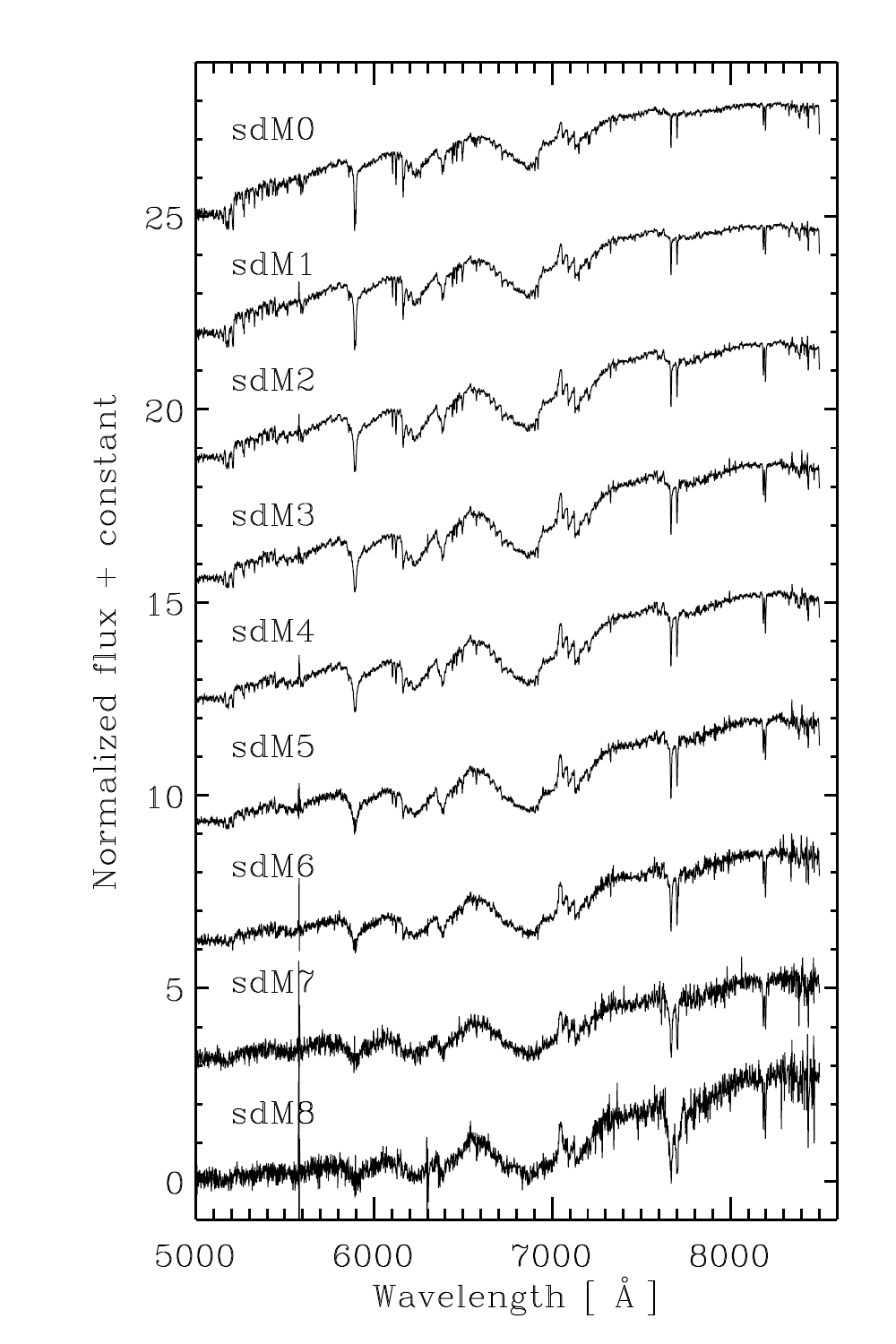}
\includegraphics[width=0.31\linewidth,angle=0]{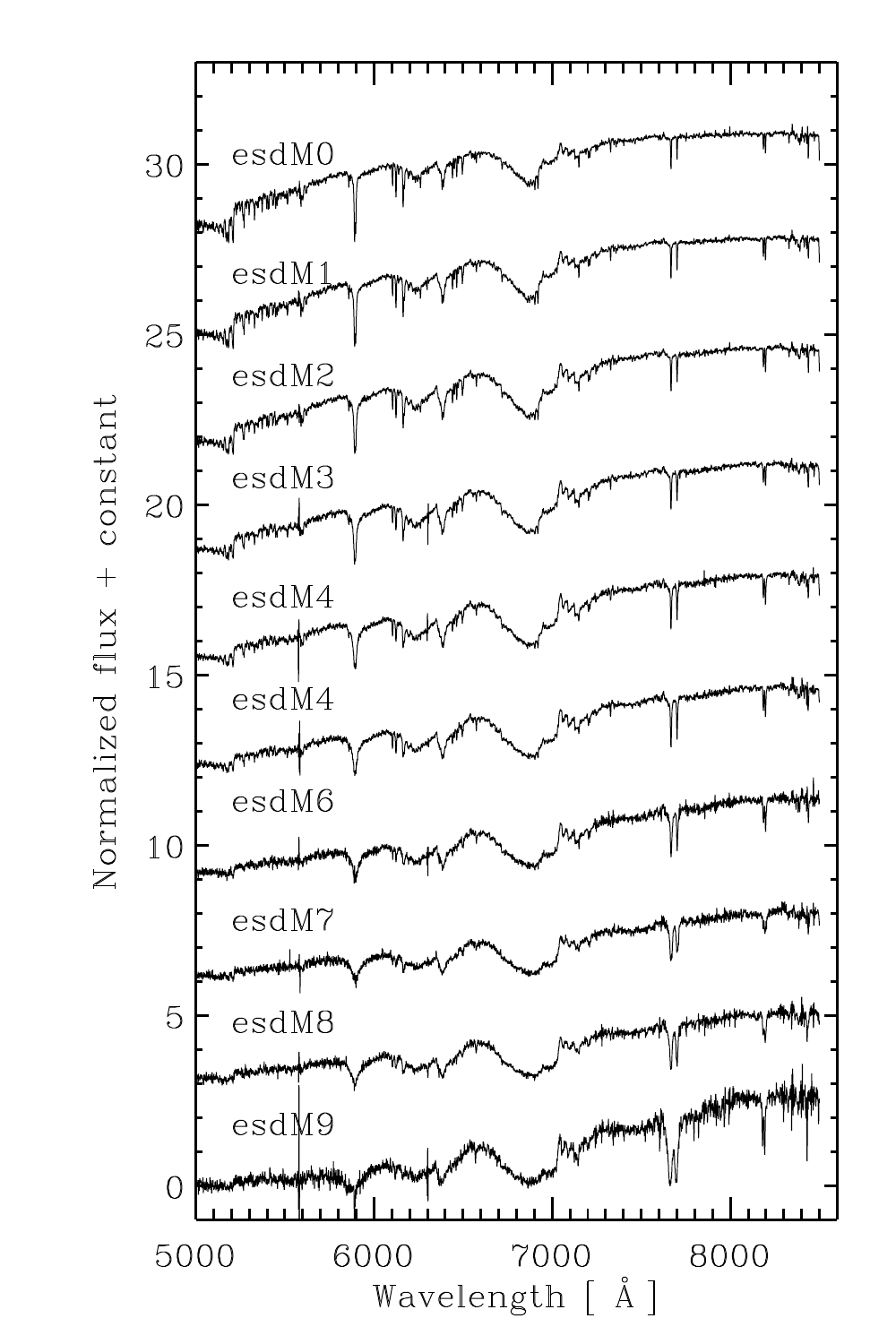}
\includegraphics[width=0.31\linewidth,angle=0]{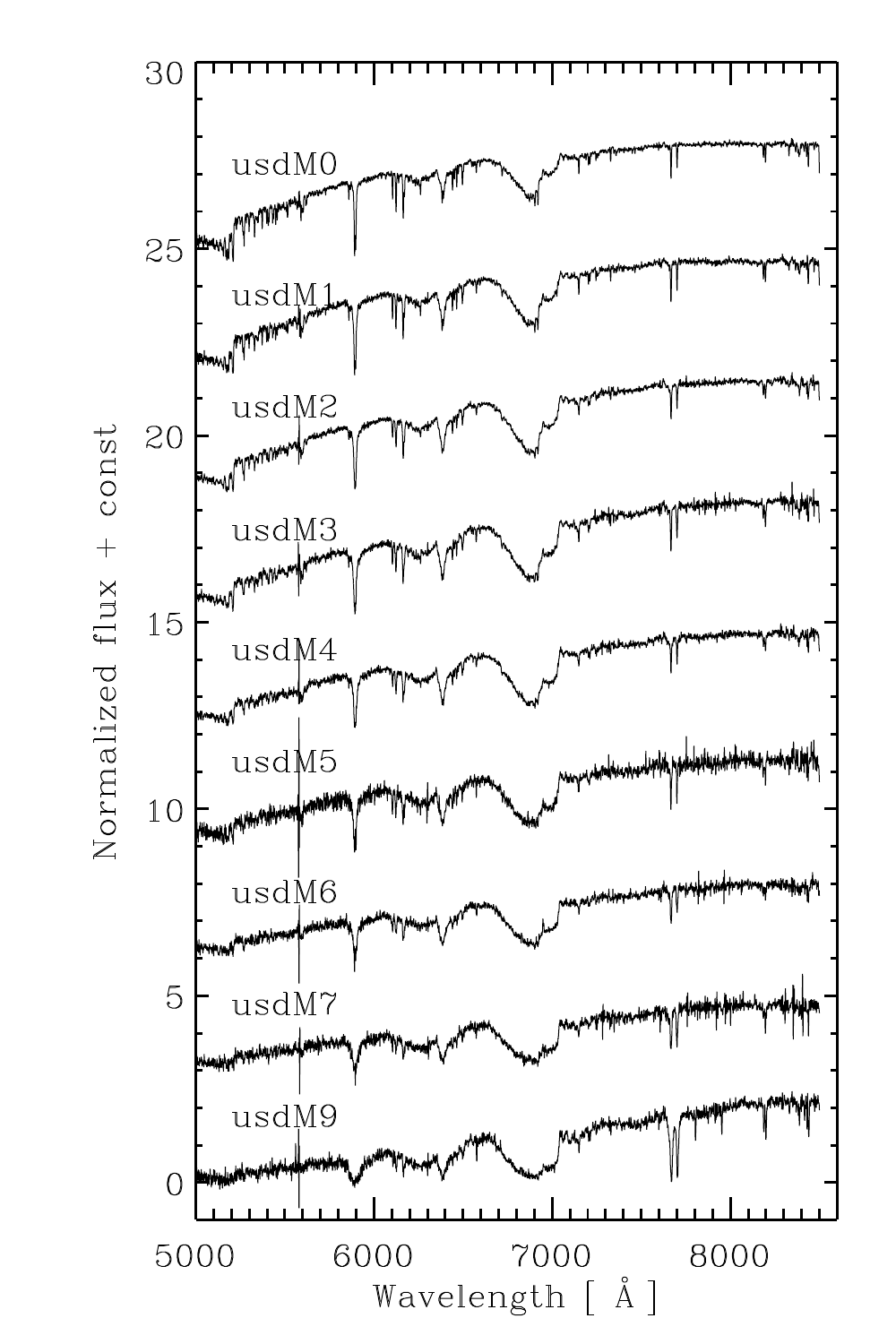}
\caption{Sequence of M-type subdwarfs (sdM; left), extreme subdwarfs (esdM, middle), and ultrasubdwarfs 
(usdM; right). Figure taken from \citet{savcheva14} with an earlier version shown in \citet{lepine07c}.}
\label{review_subBDs:fig_SpTypes}       
\end{figure}

The first spectral classification of M subdwarfs has been proposed by \citet{gizis97a}, dividing M dwarfs
into three main classes: solar-type M dwarfs with metallicity Fe/H around 0, M subdwarfs (sdM) with 
FeH\,$\sim$\,$-$1.2$\pm$0.3 dex, and extreme M subdwarfs (esdM) with approximate Fe/H of $-$2.0$\pm$0.5 dex.
This classification scheme is based on the strength of the TiO and CaH bands in the 620--740 nm
optical range. Spectral indices have been defined to infer both metal class and spectral type.
Ten years later, \citet{lepine07c} revised the boundaries of the original metal classes based on
a larger sample (factor of five bigger) of known metal-poor single and  multiple systems. They introduced a new parameter, $\tau_{TiO/CaH}$,
as well as an additional metal class with metallicities even lower than the esdM, the ultrasubdwarfs (usdM), 
to taken into account the positions of known binaries sharing the same metallicity in the (early-)M dwarf regime.
They also proposed spectral standards for each metal class and spectral types ranging from M0 to M8\@. 
The quality of the optical spectra of the sdM, esdM, and usdM templates has been improved by \citet{savcheva14} 
by stacking all SDSS spectra available per subtype and per metal class (Fig.\ \ref{review_subBDs:fig_SpTypes}). 
\citet{jao08} presented new thoughts about the naming and spectral classification of subdwarfs based 
on a sample of 88 K3--M6 sources. They propose to use the class VI of the Morgan-Keenan scheme instead 
of the terminology ``sd'' to name subdwarfs, a prefix that can be confused with the hotter sdO/sdB-type 
star which also appear sub-luminous in the HR diagram. And they showed the influence of gravity in the 
spectra of M subdwarfs and suggested to avoid classification based solely on spectral indices.

\begin{figure}
\includegraphics[width=\linewidth,angle=0]{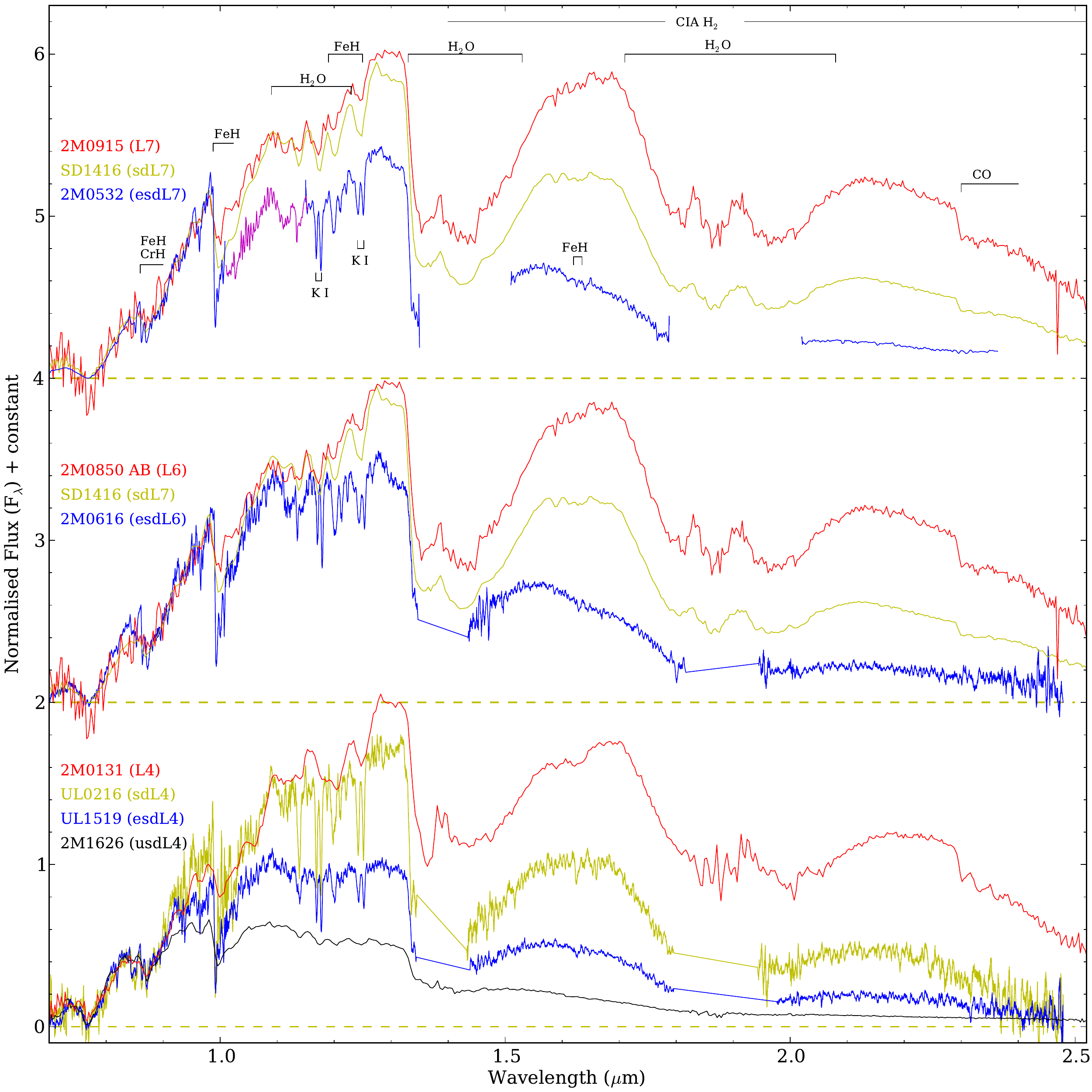}
\caption{
Optical and near-infrared spectra of L4, L6, and L7 dwarfs/subdwarfs with different subclasses. Spectra have 
been normalised at 0.89 $\mu$m. The missing wavelength region in the spectrum of 2M0532 (1.008--1.153 $\mu$m)
has been replaced by the best BT-Settl model fit in magenta (T$_{\rm eff}$\,=\,1600 K, [Fe/H]\,=\,1.6 dex, and 
$\log$(g)\,=\,5.25 dex). Figure from \citet{zhang17a}.}
\label{review_subBDs:fig_sdL}
\end{figure}

In the L dwarf, the current classification is only tentative due to the small number of L subdwarfs
announced to date and the narrow range of physical parameters. Nonetheless, several groups attempted
to extend the M dwarf classification into the L regime.
\citet{zhang17a} built on the extensions proposed by \citet{burgasser07b} and \citet{kirkpatrick16},
keeping the concept of the three metal classes proposed by \citet{lepine07c} and
applying it to the L subdwarfs (sdL, esdL, and usdL). Their spectral classification is based on
the comparison of spectra of about 30 L0--L8 subdwarfs with those of solar-type L dwarf standards
defined in the literature \citep[e.g.][]{kirkpatrick00}. They focused on key features sensitive to metallicity
and temperatures (their Table 3 and previous section) to evaluate the differences in the optical 
(CaH and VO bands around 700 nm, 
VO band at 800 nm, strength of the K{\small{I}} doublet at 770 nm, TiO band at 850 nm) and in the infrared 
(FeH band at 990 nm, CO band at 2300 nm, and the depression in the $H$ and $K$ passbands due to the 
collision-induced H$_{2}$ absorption), yielding a revised classification of known metal-poor L dwarfs
that can be used for future discoveries (Fig.\ \ref{review_subBDs:fig_sdL}).

%
%
%

%
%
%

%
%
\section{Multiplicity}
\label{review_subBDs:binary}

The multiplicity fraction and binary properties of UCSDs is poorly constrained for two main reasons.
On the one hand, most of the known UCSDs have been identified very recently and, on the other hand,
they are faint both at optical and infrared wavelengths. High-resolution imaging is feasible for a
limited sub-sample because bright reference stars are needed to close the loop as in the case of adaptive
optics for example.

Only a limited number of surveys have been conducted to look at the multiplicity of metal-poor M dwarfs over a wide range of separations.
\citet{jao09a} found that the multiplicity rate of K and M subdwarfs is slightly lower than their solar-type
counterparts (26$\pm$6\% vs 36$\pm$5\%) from an optical speckle survey. The total multiplicity fraction of
M subdwarfs can be divided up as follows: 3\% have companions within 10 au, another 3\% within the 
10--100 au range, 14\% beyond 100 au, and the remaining 6\% are spectroscopic binaries. 
Combining the outcome of a Hubble survey of 28 metal-poor M dwarfs of \citet{riaz08a} with high spatial 
lucky imaging observations of 24 M subdwarfs, \citet{lodieu09c} resolved only one system with a projected 
separation of 0.7 arcsec (LHS\,182), deriving a binary frequency of 3.7$\pm$2.6\ (1$\sigma$ confidence limit) 
for M subdwarfs (mainly M0--M5). This result is in line with the two companions exhibiting H$\alpha$
in emission mong a sample of 68 LHS objects \citep[2.9\%;][]{gizis98}.

Finally, only three M subdwarfs with metallicities below $-$0.5 dex and masses less than 0.5 M$_{\odot}$ 
have dynamical mass measurements. They belong to two double-lined eclipsing binaries with 
resolved orbits: the secondary of the $\mu$\,Cas\,AB system with a mass of 0.17 M$_{\odot}$ \citep{drummond95a}
and both components of the G\,006-026\,BC system whose masses span 0.43--0.47 M$_{\odot}$ \citep{jao16b}.

%
%
\section{Physical parameters}
\label{review_subBDs:masses}

At the time of writing, no mass estimate independent of evolutionary models exist for UCSDs because
they are either too faint or no eclipsing/spectroscopic binary exist for direct
dynamical mass measurement.

Currently, the range of physical parameters for UCSDs originates from the direct comparison of observed
optical and/or near-infrared spectra with state-of-the-art evolutionary models. \citet{burgasser08a}
identified a subdwarf with the latest spectral type reported to date (2MASS\,J0532$+$8246; sdL7), lying
at the stellar/substellar boundary.
These authors fitted the full spectral energy distribution (SED) of 2MASS\,J0532$+$8246 with the
NextGen models \citep{baraffe98}, deriving an effective temperature (T$_{\rm eff}$) of 1730$\pm$90\,K
and a mass in the range 0.0744--0.0835 M$_{\odot}$ for metallicities between 0 and $-$2.0 dex and
ages of 10--15 Gyr. However, the lithium feature at 6707.8\,\AA{} has not been detected in higher 
resolution spectra \citep{lodieu15a}, suggesting a minimum mass of 0.06 M$_{\odot}$ \citep{magazzu92,rebolo96}.
\citet{burgasser09a} repeated a similar process for a warmer L subdwarf classified as sdL4 \citep{sivarani09}
and derived a mean T$_{\rm eff}$ of 2300$\pm$200\,K, $\log$(g)\,=\,5.0--5.5 dex, and metallicity between
$-$1.0 and $-$1.5 dex using the Drift-Phoenix models \citep{helling08a}.
\citet{zhang17a} extended such a procedure to six new L subdwarfs identified in the cross-match of
SDSS and UKIDSS as well as all previously known L subdwarfs as of 2017\@. They determined T$_{\rm eff}$ 
and metallicities for 22 L subdwarfs, extreme subdwarfs, and ultrasubdwarfs with spectral types in the 
L0--L7 range by direct comparison with the BT-Settl models \citep{allard12}, yielding temperatures
of 1500--2700\,K for metallicities between $-$1.0 and $-$2.0 dex. On average, the T$_{\rm eff}$ of
subdwarfs are 100--400\,K higher than solar-metallicity L dwarfs depending on the spectral sub-type and
metal class. Some of these L subdwarfs might be brown dwarfs rather than very low-mass stellar members
of the halo based on the latest BT-Settl models \citep{zhang17b} but the exact location of the
stellar/substellar boundary at low metallicity still requires dynamical measurements.

%
%
\section{Future work}
\label{review_subBDs:future}

The past two decades have witness the existence of M, L, and T subdwarfs and the first spectral 
classification of this metal-poor population. The advent of large-scale surveys at both optical 
and infrared wavelengths (2MASS, SDSS, UKIDSS, WISE) have increased the numbers of UCSDs and 
extended the sequence to cooler metal-poor L and T dwarfs. However, our knowledge of UCSDs
still remains in its infancy. A number of improvements and discoveries are required to take 
up this field to the next level. 

\begin{itemize}
\item The improvement in the determination of physical parameters of subdwarfs requires the fitting
of optical and infrared spectral energy distributions of several hundreds of spectra to complement 
the extensive work on the properties and kinematics of subdwarfs by \citet{savcheva14}.
\item The saturation of the $\tau_{TiO/CaH}$ index around $\sim$M8--M9 suggests that a revision of the current
spectral classification scheme is needed. Moreover, a near-infrared spectral classification is also
necessary to classify future discoveries in large-scale infrared surveys.
\item The progress in the accuracy of metallicity scale for ultracool subdwarfs calls for searches of
close-in and/or wide companions to brighter subdwarfs with well-determined metallicities.
\item Searches for M, L, and T subdwarfs should be enhanced to improve the determination of the object 
density as a function of metallicity and allow for a determination of the luminosity and mass functions 
in a distance or magnitude-limited volume.
\item The discovery of short-period binaries and eclipsing binaries is heavily needed to infer 
model-independent masses over a wide range of masses and metallicities.
\end{itemize}

The future of the field sounds bright with upcoming deep photometric surveys such as the Large
Synoptic Survey Telescope \citep[LSST;][]{ivezic08a} and the Euclid mission \citep{mellier16a}
and large-scale spectroscopic campaigns planned with WHT/WEAVE \citep{dalton12} and 
VISTA/4MOST \citep{deJong14a}. Let's look forward to the first substellar subdwarfs and
long-waited mass determinations!

%
%
\bibliographystyle{spbasicHBexo}  
\bibliography{biblio_Springer_Handbook}

\begin{thebibliography}{66}
\providecommand{\natexlab}[1]{#1}
\providecommand{\url}[1]{{#1}}
\providecommand{\urlprefix}{URL }
\expandafter\ifx\csname urlstyle\endcsname\relax
  \providecommand{\doi}[1]{DOI~\discretionary{}{}{}#1}\else
  \providecommand{\doi}{DOI~\discretionary{}{}{}\begingroup
  \urlstyle{rm}\Url}\fi
\providecommand{\eprint}[2][]{\url{#2}}

\bibitem[{{Allard} et~al.(2012){Allard}, {Homeier}, and {Freytag}}]{allard12}
{Allard} F, {Homeier} D {Freytag} B (2012) {Models of very-low-mass stars,
  brown dwarfs and exoplanets}. Royal Society of London Philosophical
  Transactions Series A 370:2765--2777

\bibitem[{{Allers} and {Liu}(2013)}]{allers13}
{Allers} KN {Liu} MC (2013) {A Near-infrared Spectroscopic Study of Young Field
  Ultracool Dwarfs}. \apj 772:79

\bibitem[{{Allers} et~al.(2007){Allers}, {Jaffe}, {Luhman}, {Liu}, {Wilson},
  {Skrutskie}, {Nelson}, {Peterson}, {Smith}, and {Cushing}}]{allers07}
{Allers} KN, {Jaffe} DT, {Luhman} KL et~al. (2007) {Characterizing Young Brown
  Dwarfs Using Low-Resolution Near-Infrared Spectra}. \apj 657:511--520

\bibitem[{{Baraffe} et~al.(1998){Baraffe}, {Chabrier}, {Allard}, and
  {Hauschildt}}]{baraffe98}
{Baraffe} I, {Chabrier} G, {Allard} F {Hauschildt} PH (1998) Evolutionary
  models for solar metallicity low-mass stars: mass-magnitude relationship and
  color-magnitude diagrams 337:403--412,
  \urlprefix\url{http://adsabs.harvard.edu/cgi-bin/nph-bib_query?bibcode=1998A%26A...337..403B&db_key=AST&high=3d1b390e2520250}

\bibitem[{{Bowler} et~al.(2010){Bowler}, {Liu}, and {Dupuy}}]{bowler10a}
{Bowler} BP, {Liu} MC {Dupuy} TJ (2010) {SDSS J141624.08+134826.7: A Nearby
  Blue L Dwarf From the Sloan Digital Sky Survey}. \apj 710:45--50

\bibitem[{{Burgasser}(2004)}]{burgasser04}
{Burgasser} AJ (2004) {Discovery of a Second L Subdwarf in the Two Micron All
  Sky Survey}. \apjl 614:L73--L76

\bibitem[{{Burgasser} and {Kirkpatrick}(2006)}]{burgasser06c}
{Burgasser} AJ {Kirkpatrick} JD (2006) {Discovery of the Coolest Extreme
  Subdwarf}. \apj 645:1485--1497

\bibitem[{{Burgasser} et~al.(2002){Burgasser}, {Kirkpatrick}, {Brown}, {Reid},
  {Burrows}, {Liebert}, {Matthews}, {Gizis}, {Dahn}, {Monet}, {Cutri}, and
  {Skrutskie}}]{burgasser02}
{Burgasser} AJ, {Kirkpatrick} JD, {Brown} ME et~al. (2002) {The Spectra of T
  Dwarfs. I. Near-Infrared Data and Spectral Classification}. \apj 564:421--451

\bibitem[{{Burgasser} et~al.(2003){Burgasser}, {Kirkpatrick}, {Burrows},
  {Liebert}, {Reid}, {Gizis}, {McGovern}, {Prato}, and {McLean}}]{burgasser03b}
{Burgasser} AJ, {Kirkpatrick} JD, {Burrows} A et~al. (2003) {The First
  Substellar Subdwarf? Discovery of a Metal-poor L Dwarf with Halo Kinematics}.
  \apj 592:1186--1192

\bibitem[{{Burgasser} et~al.(2006){Burgasser}, {Geballe}, {Leggett},
  {Kirkpatrick}, and {Golimowski}}]{burgasser06a}
{Burgasser} AJ, {Geballe} TR, {Leggett} SK, {Kirkpatrick} JD {Golimowski} DA
  (2006) {A Unified Near-Infrared Spectral Classification Scheme for T Dwarfs}.
  \apj 637:1067--1093

\bibitem[{{Burgasser} et~al.(2007){Burgasser}, {Cruz}, and
  {Kirkpatrick}}]{burgasser07b}
{Burgasser} AJ, {Cruz} KL {Kirkpatrick} JD (2007) {Optical Spectroscopy of
  2MASS Color-selected Ultracool Subdwarfs}. \apj 657:494--510

\bibitem[{{Burgasser} et~al.(2008){Burgasser}, {Vrba}, {L{\'e}pine}, {Munn},
  {Luginbuhl}, {Henden}, {Guetter}, and {Canzian}}]{burgasser08a}
{Burgasser} AJ, {Vrba} FJ, {L{\'e}pine} S et~al. (2008) {Parallax and
  Luminosity Measurements of an L Subdwarf}. \apj 672:1159--1166

\bibitem[{{Burgasser} et~al.(2009){Burgasser}, {Witte}, {Helling}, {Sanderson},
  {Bochanski}, and {Hauschildt}}]{burgasser09a}
{Burgasser} AJ, {Witte} S, {Helling} C et~al. (2009) {Optical and Near-Infrared
  Spectroscopy of the L Subdwarf SDSS J125637.13-022452.4}. \apj 697:148--159

\bibitem[{{Burningham} et~al.(2013){Burningham}, {Cardoso}, {Smith}, {Leggett},
  {Smart}, {Mann}, {Dhital}, {Lucas}, {Tinney}, {Pinfield}, {Zhang}, {Morley},
  {Saumon}, {Aller}, {Littlefair}, {Homeier}, {Lodieu}, and {20
  co-authors}}]{burningham13}
{Burningham} B, {Cardoso} CV, {Smith} L et~al. (2013) {76 T dwarfs from the
  UKIDSS LAS: benchmarks, kinematics and an updated space density}. \mnras
  433:457--497

\bibitem[{{Chabrier}(2003)}]{chabrier03}
{Chabrier} G (2003) {Galactic Stellar and Substellar Initial Mass Function}.
  \pasp 115:763--795

\bibitem[{{Cruz} et~al.(2009){Cruz}, {Kirkpatrick}, and {Burgasser}}]{cruz09}
{Cruz} KL, {Kirkpatrick} JD {Burgasser} AJ (2009) {Young L Dwarfs Identified in
  the Field: A Preliminary Low-Gravity, Optical Spectral Sequence from L0 to
  L5}. \aj 137:3345--3357

\bibitem[{{Cushing} et~al.(2009){Cushing}, {Looper}, {Burgasser},
  {Kirkpatrick}, {Faherty}, {Cruz}, {Sweet}, and {Sanderson}}]{cushing09}
{Cushing} MC, {Looper} D, {Burgasser} AJ et~al. (2009) {2MASS
  J06164006$-$6407194: The First Outer Halo L Subdwarf}. \apj 696:986--993

\bibitem[{{Cushing} et~al.(2011){Cushing}, {Kirkpatrick}, {Gelino}, {Griffith},
  {Skrutskie}, {Mainzer}, {Marsh}, {Beichman}, {Burgasser}, {Prato}, {Simcoe},
  {Marley}, {Saumon}, {Freedman}, {Eisenhardt}, and {Wright}}]{cushing11}
{Cushing} MC, {Kirkpatrick} JD, {Gelino} CR et~al. (2011) {The Discovery of Y
  Dwarfs using Data from the Wide-field Infrared Survey Explorer (WISE)}. \apj
  743:50

\bibitem[{{Dalton} et~al.(2012){Dalton}, {Trager}, {Abrams}, {Carter},
  {Bonifacio}, {Aguerri}, {MacIntosh}, {Evans}, {Lewis}, {Navarro}, {Agocs},
  {Dee}, {Rousset}, {Tosh}, {Middleton}, {Pragt}, {Terrett}, {Brock}, {Benn},
  {Verheijen}, {Cano Infantes}, {Bevil}, {Steele}, {Mottram}, {Bates},
  {Gribbin}, {Rey}, {Rodriguez}, {Delgado}, {Guinouard}, {Walton}, {Irwin},
  {Jagourel}, {Stuik}, {Gerlofsma}, {Roelfsma}, {Skillen}, {Ridings},
  {Balcells}, {Daban}, {Gouvret}, {Venema}, and {Girard}}]{dalton12}
{Dalton} G, {Trager} SC, {Abrams} DC et~al. (2012) {WEAVE: the next generation
  wide-field spectroscopy facility for the William Herschel Telescope}. In:
  Ground-based and Airborne Instrumentation for Astronomy IV, \procspie, vol
  8446, p 84460P, \doi{10.1117/12.925950}

\bibitem[{{de Jong} et~al.(2014){de Jong}, {Barden}, {Bellido-Tirado},
  {Brynnel}, {Chiappini}, {Depagne}, {Haynes}, {Johl}, {Phillips}, {Schnurr},
  and {103 co-authors}}]{deJong14a}
{de Jong} RS, {Barden} S, {Bellido-Tirado} O et~al. (2014) {4MOST: 4-metre
  Multi-Object Spectroscopic Telescope}. In: Ground-based and Airborne
  Instrumentation for Astronomy V, \procspie, vol 9147, p 91470M,
  \doi{10.1117/12.2055826}

\bibitem[{{Drummond} et~al.(1995){Drummond}, {Christou}, and
  {Fugate}}]{drummond95a}
{Drummond} JD, {Christou} JC {Fugate} RQ (1995) {Full Adaptive Optics Images of
  ADS 9731 and MU Cassiopeiae: Orbits and Masses}. \apj 450:380

\bibitem[{{Geballe} et~al.(2002){Geballe}, {Knapp}, {Leggett}, {Fan},
  {Golimowski}, {Anderson}, {Brinkmann}, {Csabai}, and {21
  coauthors}}]{geballe02}
{Geballe} TR, {Knapp} GR, {Leggett} SK et~al. (2002) {Toward Spectral
  Classification of L and T Dwarfs: Infrared and Optical Spectroscopy and
  Analysis}. \apj 564:466--481

\bibitem[{{Gizis}(1997)}]{gizis97a}
{Gizis} JE (1997) {M-Subdwarfs: Spectroscopic Classification and the
  Metallicity Scale}. \aj 113:806--822

\bibitem[{{Gizis}(1998)}]{gizis98}
{Gizis} JE (1998) {High Chromospheric Activity in M Subdwarfs}. \aj
  115:2053--2058

\bibitem[{{Gizis} and {Reid}(1997)}]{gizis97b}
{Gizis} JE {Reid} IN (1997) {Probing the LHS Catalog. I. New Nearby Stars and
  the Coolest Subdwarf}. \pasp 109:849--856

\bibitem[{{Gizis} et~al.(1997){Gizis}, {Scholz}, {Irwin}, and
  {Jahreiss}}]{gizis97c}
{Gizis} JE, {Scholz} RD, {Irwin} M {Jahreiss} H (1997) {APMPM J1523-0245 - A
  new high proper motion cool subdwarf}. \mnras 292:L41--L43

\bibitem[{{Helling} et~al.(2008){Helling}, {Ackerman}, {Allard}, {Dehn},
  {Hauschildt}, {Homeier}, {Lodders}, {Marley}, {Rietmeijer}, {Tsuji}, and
  {Woitke}}]{helling08a}
{Helling} C, {Ackerman} A, {Allard} F et~al. (2008) {A comparison of chemistry
  and dust cloud formation in ultracool dwarf model atmospheres}. \mnras
  391:1854--1873

\bibitem[{{Ivezic} et~al.(2008){Ivezic}, {Axelrod}, {Brandt}, {Burke},
  {Claver}, {Connolly}, {Cook}, {Gee}, {Gilmore}, {Jacoby}, {Jones}, {Kahn},
  {Kantor}, {Krabbendam}, {Lupton}, {Monet}, {Pinto}, {Saha}, {Schalk},
  {Schneider}, {Strauss}, {Stubbs}, {Sweeney}, {Szalay}, {Thaler}, {Tyson}, and
  {LSST Collaboration}}]{ivezic08a}
{Ivezic} Z, {Axelrod} T, {Brandt} WN et~al. (2008) {Large Synoptic Survey
  Telescope: From Science Drivers To Reference Design}. Serbian Astronomical
  Journal 176:1--13

\bibitem[{{Jao} et~al.(2008){Jao}, {Henry}, {Beaulieu}, and
  {Subasavage}}]{jao08}
{Jao} WC, {Henry} TJ, {Beaulieu} TD {Subasavage} JP (2008) {Cool Subdwarf
  Investigations. I. New Thoughts on the Spectral Types of K and M Subdwarfs}.
  \aj 136:840--880

\bibitem[{{Jao} et~al.(2009){Jao}, {Mason}, {Hartkopf}, {Henry}, and
  {Ramos}}]{jao09a}
{Jao} WC, {Mason} BD, {Hartkopf} WI, {Henry} TJ {Ramos} SN (2009) {Cool
  Subdwarf Investigations. II. Multiplicity}. \aj 137:3800--3808

\bibitem[{{Jao} et~al.(2016){Jao}, {Nelan}, {Henry}, {Franz}, and
  {Wasserman}}]{jao16b}
{Jao} WC, {Nelan} EP, {Henry} TJ, {Franz} OG {Wasserman} LH (2016) {Cool
  Subdwarf Investigations. III. Dynamical Masses of Low-metallicity Subdwarfs}.
  \aj 152:153

\bibitem[{{Joy}(1947)}]{joy47a}
{Joy} AH (1947) {Radial Velocities and Spectral Types of 181 Dwarf Stars.} \apj
  105:96

\bibitem[{{Kirkpatrick} et~al.(1999){Kirkpatrick}, {Reid}, {Liebert}, {Cutri},
  {Nelson}, {Beichman}, {Dahn}, {Monet}, {Gizis}, and
  {Skrutskie}}]{kirkpatrick99}
{Kirkpatrick} JD, {Reid} IN, {Liebert} J et~al. (1999) {Dwarfs Cooler than
  ``M'': The Definition of Spectral Type ``L'' Using Discoveries from the 2
  Micron All-Sky Survey (2MASS)}. \apj 519:802--833,
  \urlprefix\url{http://adsabs.harvard.edu/cgi-bin/nph-bib_query?bibcode=1999ApJ...519..802K&db_key=AST}

\bibitem[{{Kirkpatrick} et~al.(2000){Kirkpatrick}, {Reid}, {Liebert}, {Gizis},
  {Burgasser}, {Monet}, {Dahn}, {Nelson}, and {Williams}}]{kirkpatrick00}
{Kirkpatrick} JD, {Reid} IN, {Liebert} J et~al. (2000) {67 Additional L Dwarfs
  Discovered by the Two Micron All Sky Survey}. \aj 120:447--472,
  \urlprefix\url{http://adsabs.harvard.edu/cgi-bin/nph-bib_query?bibcode=2000AJ....120..447K&db_key=AST}

\bibitem[{{Kirkpatrick} et~al.(2012){Kirkpatrick}, {Gelino}, {Cushing}, {Mace},
  {Griffith}, {Skrutskie}, {Marsh}, {Wright}, {Eisenhardt}, {McLean},
  {Mainzer}, {Burgasser}, {Tinney}, {Parker}, and {Salter}}]{kirkpatrick12}
{Kirkpatrick} JD, {Gelino} CR, {Cushing} MC et~al. (2012) {Further Defining
  Spectral Type ''Y'' and Exploring the Low-mass End of the Field Brown Dwarf
  Mass Function}. \apj 753:156

\bibitem[{{Kirkpatrick} et~al.(2014){Kirkpatrick}, {Schneider},
  {F{\aj}ardo-Acosta}, {Gelino}, {Mace}, {Wright}, {Logsdon}, {McLean},
  {Cushing}, {Skrutskie}, {Eisenhardt}, {Stern}, {Balokovi{\'c}}, {Burgasser},
  {Faherty}, {Lansbury}, {Rich}, {Skrzypek}, {Fowler}, {Cutri}, {Masci},
  {Conrow}, {Grillmair}, {McCallon}, {Beichman}, and {Marsh}}]{kirkpatrick14}
{Kirkpatrick} JD, {Schneider} A, {F{\aj}ardo-Acosta} S et~al. (2014) {The
  AllWISE Motion Survey and the Quest for Cold Subdwarfs}. \apj 783:122

\bibitem[{{Kirkpatrick} et~al.(2016){Kirkpatrick}, {Kellogg}, {Schneider},
  {Fajardo-Acosta}, {Cushing}, {Greco}, {Mace}, {Gelino}, {Wright},
  {Eisenhardt}, {Stern}, {Faherty}, {Sheppard}, {Lansbury}, {Logsdon},
  {Martin}, {McLean}, {Schurr}, {Cutri}, and {Conrow}}]{kirkpatrick16}
{Kirkpatrick} JD, {Kellogg} K, {Schneider} AC et~al. (2016) {The AllWISE Motion
  Survey, Part 2}. \apjs 224:36

\bibitem[{{Leggett} et~al.(2000){Leggett}, {Geballe}, {Fan}, {Schneider},
  {Gunn}, {Lupton}, {Knapp}, {Strauss}, and {27 coauthors}}]{leggett00}
{Leggett} SK, {Geballe} TR, {Fan} X et~al. (2000) {The Missing Link: Early
  Methane (``T'') Dwarfs in the Sloan Digital Sky Survey}. \apjl 536:L35--L38

\bibitem[{{L{\'e}pine} and {Scholz}(2008)}]{lepine08b}
{L{\'e}pine} S {Scholz} RD (2008) {Twenty-Three New Ultracool Subdwarfs from
  the Sloan Digital Sky Survey}. \apjl 681:L33--L36

\bibitem[{{L{\'e}pine} et~al.(2003){L{\'e}pine}, {Shara}, and
  {Rich}}]{lepine03d}
{L{\'e}pine} S, {Shara} MM {Rich} RM (2003) {Discovery of an Ultracool
  Subdwarf: LSR 1425+7102, the First Star with Spectral Type sdM8.0}. \apjl
  585:L69--L72

\bibitem[{{L{\'e}pine} et~al.(2007){L{\'e}pine}, {Rich}, and
  {Shara}}]{lepine07c}
{L{\'e}pine} S, {Rich} RM {Shara} MM (2007) {Revised Metallicity Classes for
  Low-Mass Stars: Dwarfs (dM), Subdwarfs (sdM), Extreme Subdwarfs (esdM), and
  Ultrasubdwarfs (usdM)}. \apj 669:1235--1247

\bibitem[{{Lodieu} et~al.(2005){Lodieu}, {Scholz}, {McCaughrean}, {Ibata},
  {Irwin}, and {Zinnecker}}]{lodieu05b}
{Lodieu} N, {Scholz} RD, {McCaughrean} MJ et~al. (2005) {Spectroscopic
  classification of red high proper motion objects in the Southern Sky}. \aap
  440:1061--1078

\bibitem[{{Lodieu} et~al.(2009){Lodieu}, {Zapatero Osorio}, and
  {Mart{\'{\i}}n}}]{lodieu09c}
{Lodieu} N, {Zapatero Osorio} MR {Mart{\'{\i}}n} EL (2009) {Lucky Imaging of M
  subdwarfs}. \aap 499:729--736

\bibitem[{{Lodieu} et~al.(2010){Lodieu}, {Zapatero Osorio}, {Mart{\'{\i}}n},
  {Solano}, and {Aberasturi}}]{lodieu10a}
{Lodieu} N, {Zapatero Osorio} MR, {Mart{\'{\i}}n} EL, {Solano} E {Aberasturi} M
  (2010) {GTC/OSIRIS Spectroscopic Identification of a Faint L Subdwarf in the
  UKIRT Infrared Deep Sky Survey}. \apjl 708:L107--L111

\bibitem[{{Lodieu} et~al.(2012){Lodieu}, {Espinoza Contreras}, {Zapatero
  Osorio}, {Solano}, {Aberasturi}, and {Mart{\'{\i}}n}}]{lodieu12b}
{Lodieu} N, {Espinoza Contreras} M, {Zapatero Osorio} MR et~al. (2012) {New
  ultracool subdwarfs identified in large-scale surveys using Virtual
  Observatory tools. I. UKIDSS LAS DR5 vs. SDSS DR7}. \aap 542:A105

\bibitem[{{Lodieu} et~al.(2015){Lodieu}, {Burgasser}, {Pavlenko}, and
  {Rebolo}}]{lodieu15a}
{Lodieu} N, {Burgasser} AJ, {Pavlenko} Y {Rebolo} R (2015) {A search for
  lithium in metal-poor L dwarfs}. \aap 579:A58

\bibitem[{{Lodieu} et~al.(2017){Lodieu}, {Espinoza Contreras}, {Zapatero
  Osorio}, {Solano}, {Aberasturi}, {Mart{\'{\i}}n}, and {Rodrigo}}]{lodieu17a}
{Lodieu} N, {Espinoza Contreras} M, {Zapatero Osorio} MR et~al. (2017) {New
  ultracool subdwarfs identified in large-scale surveys using Virtual
  Observatory tools}. \aap 598:A92

\bibitem[{{Mace} et~al.(2013){Mace}, {Kirkpatrick}, {Cushing}, {Gelino},
  {McLean}, {Logsdon}, {Wright}, {Skrutskie}, {Beichman}, {Eisenhardt}, and
  {Kulas}}]{mace13b}
{Mace} GN, {Kirkpatrick} JD, {Cushing} MC et~al. (2013) {The Exemplar T8
  Subdwarf Companion of Wolf 1130}. \apj 777:36

\bibitem[{{Magazzu} et~al.(1992){Magazzu}, {Rebolo}, and
  {Pavlenko}}]{magazzu92}
{Magazzu} A, {Rebolo} R {Pavlenko} IV (1992) {Lithium abundances in classical
  and weak T Tauri stars}. \apj 392:159--171

\bibitem[{{Mart\'{\i}n} et~al.(1999){Mart\'{\i}n}, {Delfosse}, {Basri},
  {Goldman}, {Forveille}, and {Zapatero Osorio}}]{martin99a}
{Mart\'{\i}n} EL, {Delfosse} X, {Basri} G et~al. (1999) {Spectroscopic
  Classification of Late-M and L Field Dwarfs}. \aj 118:2466--2482,
  \urlprefix\url{http://adsabs.harvard.edu/cgi-bin/nph-bib_query?bibcode=1999AJ....118.2466M&db_key=AST}

\bibitem[{{Mellier}(2016)}]{mellier16a}
{Mellier} Y (2016) {Euclid and the Dark Universe}. In: 41st COSPAR Scientific
  Assembly, abstracts from the meeting that was to be held 30 July - 7 August
  at the Istanbul Congress Center (ICC), Turkey, but was cancelled. See <A
  href=''http://cospar2016.tubitak.gov.tr/en/''>http://cospar2016.tubitak.gov.tr/en/</A>,
  Abstract H0.2-1-16., COSPAR Meeting, vol~41

\bibitem[{{Morgan} et~al.(1943){Morgan}, {Keenan}, and {Kellman}}]{morgan43}
{Morgan} WW, {Keenan} PC {Kellman} E (1943) {An atlas of stellar spectra, with
  an outline of spectral classification}. Chicago, Ill., The University of
  Chicago press

\bibitem[{{Pinfield} et~al.(2012){Pinfield}, {Burningham}, {Lodieu}, {Leggett},
  {Tinney}, {van Spaandonk}, {Marocco}, {Smart}, {Gomes}, {Smith}, {Lucas},
  {Day-Jones}, {Murray}, {Katsiyannis}, {Catalan}, {Cardoso}, {Clarke},
  {Folkes}, {G{\'a}lvez-Ortiz}, {Homeier}, {Jenkins}, {Jones}, and
  {Zhang}}]{pinfield12}
{Pinfield} DJ, {Burningham} B, {Lodieu} N et~al. (2012) {Discovery of the
  benchmark metal-poor T8 dwarf BD +01{$\deg$}2920B}. \mnras 422:1922--1932

\bibitem[{{Rebolo} et~al.(1996){Rebolo}, {Mart\'{\i}n}, {Basri}, {Marcy}, and
  {Zapatero-Osorio}}]{rebolo96}
{Rebolo} R, {Mart\'{\i}n} EL, {Basri} G, {Marcy} GW {Zapatero-Osorio} MR (1996)
  {Brown Dwarfs in the Pleiades Cluster Confirmed by the Lithium Test}. \apjl
  469:L53

\bibitem[{{Reid} and {Hawley}(2005)}]{reid05a}
{Reid} IN {Hawley} SL (2005) {New light on dark stars : red dwarfs, low-mass
  stars, brown dwarfs}. \doi{10.1007/3-540-27610-6}

\bibitem[{{Riaz} et~al.(2008){Riaz}, {Gizis}, and {Samaddar}}]{riaz08a}
{Riaz} B, {Gizis} JE {Samaddar} D (2008) {Hubble Space Telescope Search for M
  Subdwarf Binaries}. \apj 672:1153--1158

\bibitem[{{Savcheva} et~al.(2014){Savcheva}, {West}, and
  {Bochanski}}]{savcheva14}
{Savcheva} AS, {West} AA {Bochanski} JJ (2014) {A New Sample of Cool Subdwarfs
  from SDSS: Properties and Kinematics}. \apj 794:145

\bibitem[{{Schmidt}(1975)}]{schmidt75a}
{Schmidt} M (1975) {The mass of the galactic halo derived from the luminosity
  function of high-velocity stars}. \apj 202:22--29

\bibitem[{{Schmidt} et~al.(2010){Schmidt}, {West}, {Burgasser}, {Bochanski},
  and {Hawley}}]{schmidt10a}
{Schmidt} SJ, {West} AA, {Burgasser} AJ, {Bochanski} JJ {Hawley} SL (2010)
  {Discovery of an Unusually Blue L Dwarf Within 10 pc of the Sun}. \aj
  139:1045--1050

\bibitem[{{Scholz} et~al.(2004{\natexlab{a}}){Scholz}, {Lehmann}, {Matute}, and
  {Zinnecker}}]{scholz04c}
{Scholz} R, {Lehmann} I, {Matute} I {Zinnecker} H (2004{\natexlab{a}}) {The
  nearest cool white dwarf (d$\sim$4 pc), the coolest M-type subdwarf (sdM9.5),
  and other high proper motion discoveries}. \aap 425:519--527

\bibitem[{{Scholz} et~al.(2004{\natexlab{b}}){Scholz}, {Lodieu}, and
  {McCaughrean}}]{scholz04b}
{Scholz} RD, {Lodieu} N {McCaughrean} MJ (2004{\natexlab{b}}) {SSSPM
  J1444-2019: An extremely high proper motion, ultracool subdwarf}. \aap
  428:L25--L28

\bibitem[{{Schweitzer} et~al.(1999){Schweitzer}, {Scholz}, {Stauffer}, {Irwin},
  and {McCaughrean}}]{schweitzer99}
{Schweitzer} A, {Scholz} RD, {Stauffer} J, {Irwin} M {McCaughrean} MJ (1999)
  {APMPM J0559-2903: The coolest extreme subdwarf known}. \aap 350:L62--L64

\bibitem[{{Sivarani} et~al.(2009){Sivarani}, {L{\'e}pine}, {Kembhavi}, and
  {Gupchup}}]{sivarani09}
{Sivarani} T, {L{\'e}pine} S, {Kembhavi} AK {Gupchup} J (2009) {SDSS
  J125637-022452: A High Proper Motion L Subdwarf}. \apjl 694:L140--L143

\bibitem[{{Zhang} et~al.(2013){Zhang}, {Pinfield}, {Burningham}, {Jones},
  {G{\'a}lvez-Ortiz}, {Catal{\'a}n}, {Smart}, {L{\'e}pine}, {Clarke},
  {Pavlenko}, {Murray}, {Kuznetsov}, {Day-Jones}, {Gomes}, {Marocco}, and
  {Sip{\"o}cz}}]{zhang13}
{Zhang} ZH, {Pinfield} DJ, {Burningham} B et~al. (2013) {A spectroscopic and
  proper motion search of Sloan Digital Sky Survey: red subdwarfs in binary
  systems}. \mnras 434:1005--1027

\bibitem[{{Zhang} et~al.(2017{\natexlab{a}}){Zhang}, {Homeier}, {Pinfield},
  {Lodieu}, {Jones}, {Allard}, and {Pavlenko}}]{zhang17b}
{Zhang} ZH, {Homeier} D, {Pinfield} DJ et~al. (2017{\natexlab{a}}) {Primeval
  very low-mass stars and brown dwarfs --- II. The most metal-poor substellar
  object}. \mnras 468:261--271

\bibitem[{{Zhang} et~al.(2017{\natexlab{b}}){Zhang}, {Pinfield},
  {G{\'a}lvez-Ortiz}, {Burningham}, {Lodieu}, {Marocco}, {Burgasser},
  {Day-Jones}, {Allard}, {Jones}, {Homeier}, {Gomes}, and {Smart}}]{zhang17a}
{Zhang} ZH, {Pinfield} DJ, {G{\'a}lvez-Ortiz} MC et~al. (2017{\natexlab{b}})
  {Primeval very low-mass stars and brown dwarfs - I. Six new L subdwarfs,
  classification and atmospheric properties}. \mnras 464:3040--3059

\end{thebibliography}

%
%
\begin{acknowledgement}
NL is supported by programme AYA2015-69350-C3-2-P from Spanish Ministry of Economy and Competitiveness (MINECO).
NL thanks ZengHua Zhang for his input on the review.
\end{acknowledgement}
\end{document}